\documentclass{iaus}
\usepackage{graphicx}
\newcommand{\fdg}{\mbox{\ensuremath{.\!\!^\circ}}}
\newcommand{\arcdeg}{\ensuremath{^{\circ}}}

\title[Magellanic Clouds Chemical Enrichment History] 
{The Chemical Enrichment History of the Magellanic Clouds Field Populations}

\author[Ricardo Carrera \etal\ ]   
{Ricardo Carrera$^1$
 Carme Gallart$^2$
Antonio Aparicio$^2$
Edgardo Costa$^3$
Eduardo Hardy$^{3,4}$
Rene A. M\'endez$^3$
Noelia E. D. N\"oel$^2$
Robert Zinn$^5$}

\affiliation{$^1$Osservatorio Astronomico di Bologna, Via Ranzani 1, I-40127 Bologna, Italy 
\\ email: {\tt ricardo.carrera@oabo.inaf.it} \\[\affilskip]
$^2$Instituto de Astrof\'{\i}sica de Canarias, V\'{\i}a L\'actea sn, E-38200 La Laguna, Spain 
\\[\affilskip]
$^3$Departamento de Astronom\'{\i}a, Universidad de Chile, Casilla 36--D, Santiago, Chile
\\[\affilskip]
$^4$National Radio Astronomy Observatory, Casilla El Golf 16-10, Las Condes, Santiago, Chile
\\[\affilskip]
$^5$Department of Astronomy, Yale University, New Heaven, USA}

\pubyear{2008}
\volume{256}  
\pagerange{119--126}
\jname{The Magellanic System: Stars, Gas, and Galaxies}
\editors{Jacco Th. van Loon \& Joana M. Oliveira, eds.}
\begin{document}

\maketitle

\begin{abstract}

We report the results of our project devoted to study the chemical
enrichment history of the field population in the Magellanic Clouds using
Ca II triplet spectroscopy. 

\end{abstract}

\firstsection 
\section{Introduction}

The study of the ages and metallicities of the resolved stars in a nearby
galaxy provides very detailed information on its evolutionary history. The
Magellanic Clouds are examples of galaxies where this method can be
applied very successfully. However, their vastness and our limitations in
observing sizable samples of stars imply that there are particularly large
gaps of knowledge in this area, compared with others. For example, their
chemical enrichment histories have mainly been studied from their cluster
systems. However, these objects have some drawbacks, like the age-gap of
the LMC clusters or the single old cluster known in the SMC. Although the
SMC is more metal-poor than the LMC, the chemical evolution of the cluster
systems has been qualitatively similar in both galaxies. They show a first
episode of chemical enrichment followed by a period of slow chemical
evolution until around 4 Gyrs ago. Then, both galaxies took off again
(\cite[Olszewski \etal\ 1991]{ol91};\cite[Piatti \etal\ 2001]{piatti01}). However, as mentioned above, there are some epochs in which the
lack of clusters makes it difficult to extract definite conclusions. The
chemical evolution of the Magellanic Clouds has also been investigated
from their planetary nebulae (\cite[Dopita \etal\ 1991]{dopita97};\cite[Idiart \etal\ 2007]{idiart07}), which show a behaviour similar to the
clusters one. Finally, \cite[Cole \etal\ (2005)]{c05} have studied the
chemical evolution of the LMC bar based on a sample of red giant branch
(RGB) stars. The age-metallicity relationship (AMR) of the bar stars is
similar to the cluster's one, particularly for the older ages. However,
the increase of metallicity observed in the clusters in the last 3 Gyrs is
not observed in the bar.

What about the field populations in the LMC disk and in the SMC? Do they
share the cluster and planetary nebulae behaviour in each galaxy? Does
chemical evolution show a global pattern in each galaxy or on the
contrary, does it depends on the position? To address these questions, and
as part of a large project devoted to study the stellar content of the
Magellanic Clouds, we have obtained metallicities and ages for stars in 4
large LMC and 13 SMC fields. In this paper we will discuss the procedure
used to derive metallicities and ages, and describe our main conclusions
on the chemical enrichment history of both Magellanic Clouds.

\section{Deriving Stellar Metallicities and Ages}\label{sec_cat}

Although the best way to derive chemical abundances is high-resolution
spectroscopy, this technique needs huge amounts of telescope time to
measure a significant number of stars even in the nearest galaxies. The
alternative is low-resolution spectroscopy which, together with the modern
multi-object spectrographs, allows us to observe large samples in a
reasonable time. However, in external galaxies, even low-resolution
spectroscopy is only possible for the brightest objects, which in most 
cases are RGB stars. A powerfull index to derive metallicities in these
stars is the infrared Ca II triplet (CaT). This index is linearly correlated
with metallicity in the range $-2.2\leq$[Fe/H]$\leq$+0.47, with no
measurable influence of age in the interval 13$\leq$Age/Gyr$\leq$0.25
\cite[(Carrera \etal\ 2007)]{carrera07}. Tough this index may overestimate
metallicities for values below -2.5 dex \cite[(Koch \etal\ 2008)]{koch08},
the range of ages and metallicities in which it shows a linear behaviour
agrees with the ones expected in the Magellanic Clouds.

The position of the RGB on the color-magnitude diagram (CMD) suffers from
age--metallicity degeneracy. However, when the metallicity has been
obtained in an alternative way, as in this case from spectroscopy, this
age--metallicity degeneracy can be broken up, and stellar ages can be
estimated from the position of the stars in the CMD. To do that, we
computed a polynomial relationship to derive stellar ages from
metallicities and positions of stars in the CMD. For that purpose, a
synthetic CMD computed with IAC-STAR \cite[(Aparicio \& Gallart
2004)]{aparicio04} with the overshooting BaSTI stellar stellar evolution
models \cite[(Pietrinferni \etal\ 2004)]{pie04} as input was used. Details
of this procedure can be found in \cite [Carrera \etal\
(2008a)]{carrera08a} and \cite[Carrera \etal\ (2008b)]{carrera08b}.

\section{LMC Chemical Enrichment History}\label{sec_lmc}

In the LMC, we have studied four fields located at 2\fdg3, 4\fdg 0, 5\fdg
5 and 7\fdg1 northward of the bar. The CMDs of these fields have been
presented by \cite[Gallart \etal\ (2008)]{gallart08}. We are obtaining
detailed SFHs for them (see Gallart \etal\ 2008 and \cite [Meschin \etal\
2008]{meschin08} in these proceedings) using the synthetic CMD technique.
In each field we observed spectroscopically more than 100 stars in the
upper part of the RGB. The results of this work can be found in
\cite[Carrera \etal\ (2008a)]{carrera08a}. Table \,\ref{tab1} summarizes
the mean metallicities and their dispersions in each field. The inner
fields have a similar mean metallicity, while it is a factor of two more
metal-poor in the outermost field.

\begin{table}[h!]
  \begin{center}
  \caption{Mean values of metallicity distributions of LMC fields.}
  \label{tab1}
 {\small
  \begin{tabular}{|l|c|c|}\hline 
Field & $\langle$[Fe/H]$\rangle$ & $\sigma_{[Fe/H]}$\\ \hline
Bar & -0.39 & 0.19 \\
2\fdg3 & -0.47 & 0.31 \\
4\fdg0 & -0.50 & 0.37 \\
5\fdg5 & -0.45 & 0.31 \\
7\fdg1 & -0.79 & 0.44  \\ \hline
  \end{tabular}
  }
 \end{center}
\end{table}

\begin{figure}[h!]
\begin{center} \includegraphics[width=4.3in]{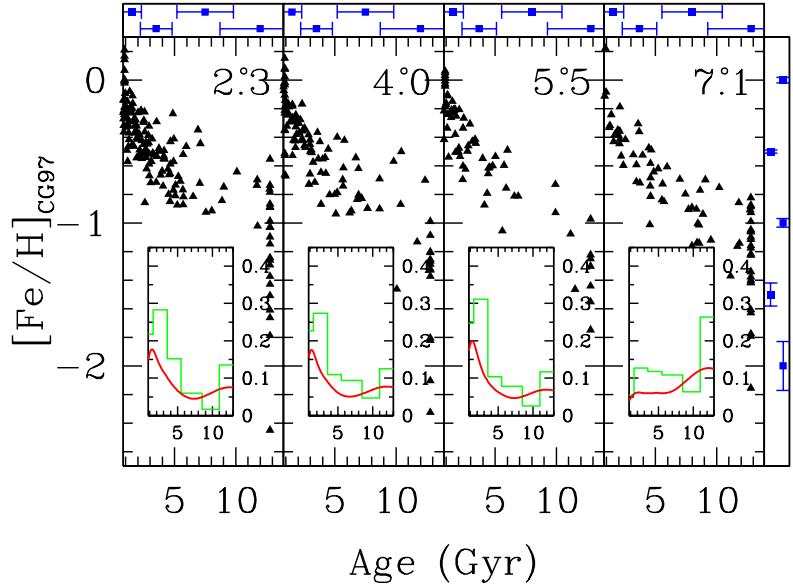} 
 \caption{Age--metallicity relationships for the four LMC fields in our sample. Inset panels show the age
distribution computed taking into account (\textsl{solid line}) and not (\textsl{histogram}) the age determination uncertainties.
The top panel show the age error in each interval. The left panels show the metallicity
error.}
\label{fig2}
\end{center}
\end{figure}

To understand why the outermost field is more metal-poor, we obtained the AMR
of each field (Fig.\,\ref{fig2}). The age error in each age interval is
indicated in the top panel. The age distributions for each field have been
plotted in inset panels taking into account (\textsl{solid line}) and not
(\textsl{histogram}) the age determination uncertainties (see
\cite[Carrera \etal\ 2008a]{carrera08a} for details).

The AMR is, within the uncertainties, very similar for all fields, and
also similar to the clusters one. As expected, the most metal-poor stars
in each field are also the oldest  ones. A rapid chemical enrichment at a
very early epoch is followed by a period of very slow metallicity
evolution until around 3 Gyr ago, when the galaxy started another period
of chemical enrichment that is still ongoing. Furthermore, the age
histograms for the three innermost fields are similar, although the total
number of stars decreases when moving away from the centre. The outermost
field has a lower fraction of ''young'' (1--4 Gyr) intermediate-age stars.
This indicates that its lower mean metallicity is related to the lower
fraction of intermediate-age, more metal-rich stars rather than to a
different chemical enrichment history (for example, a slower metal
enrichment).

\section{SMC Chemical Enrichment History}\label{sec_smc}

\begin{table}[h!]
  \begin{center}
  \caption{SMC observed fields, mean metallicities and dispersion}
  \label{tab2}
 {\footnotesize
  \begin{tabular}{|l|c|c|c|c|c|c|c|}\hline 
Field & $\alpha_{2000}$ & $\delta_{2000}$ & r(') & PA (\arcdeg) &  Zone & $\langle$[Fe/H]$\rangle$ & $\sigma_{[Fe/H]}$\\ \hline
{\sl smc0057} & {\sl 00:57} & {\sl -73:53} & {\sl 65.7} & {\sl 164.4} & {\sl South}  & {\sl -1.01} & {\sl 0.33}\\
{\bf qj0037} & {\bf 00:37} & {\bf -72:18} & {\bf 78.5} & {\bf 294.0} & {\bf West}  & {\bf -0.95} & {\bf 0.17}\\
{\bf qj0036} & {\bf 00:36} & {\bf -72:25} & {\bf 79.8} & {\bf 288.0} & {\bf West}& {\bf -0.98} & {\bf 0.25}\\
{\emph qj0111} & {\emph 01:11} & {\emph -72:49} & {\emph 80.9} & {\emph 89.5} & {\emph East} & {\emph -1.08} & {\emph 0.21}\\
{\emph qj0112} & {\emph 01:12} & {\emph -72:36} & {\emph 87.4} & {\emph 81.0 }& {\emph East} & {\emph -1.16} & {\emph 0.32}\\
{\bf qj0035} & {\bf 00:35} & {\bf -72:01} & {\bf 95.5} & {\bf 300.6} &{\bf  West} & {\bf -1.09} & {\bf 0.24}\\
{\emph qj0116} & {\emph 01:16} & {\emph -72:59} & {\emph 102.5} & {\emph 95.2} & {\emph East} & {\emph -0.96} & {\emph 0.26}\\
{\sl smc0100} & {\sl 01:00} & {\sl -74:57} & {\sl 130.4} & {\sl 167.5} & {\sl South} & {\sl -1.07} & {\sl 0.28}\\
{\sl qj0047} & {\sl 00:47} & {\sl -75:30} & {\sl 161.7} & {\sl 187.7} & {\sl South} & {\sl -1.15} & {\sl 0.27}\\
{\bf qj0033} & {\bf 00:33} & {\bf -70:28} & {\bf 172.9} & {\bf 325.0} & {\bf West} & {\bf -1.58} & {\bf 0.57}\\
{\sl smc0049} & {\sl 00:49} & {\sl -75:44} & {\sl 174.8} & {\sl 184.6} & {\sl South} & {\sl -1.00} & {\sl 0.28} \\
{\sl qj0102} & {\sl 01:02} & {\sl -74:46} & {\sl 179.5} & {\sl 169.4} & {\sl South} & {\sl -1.29} & {\sl 0.42}\\
{\sl smc0053} & {\sl 00:53} & {\sl -76:46} & {\sl 236.3} & {\sl 179.4} & {\sl South} & {\sl -1.64} & {\sl 0.50}\\ \hline
  \end{tabular}
  }
 \end{center}
\end{table}

In the SMC we secured spectroscopy of stars in 13 fields spread about the
galaxy body. The corresponding photometry has been published by 
\cite[No\"el \etal\ (2007)]{noel07} and their detailed SFHs have been
presented in this volume by \cite[No\"el \etal\ (2008)]{noel08}. The
positions of these fields, together with their mean metallicities are
listed in Table \,\ref{tab2}. The fields are ordered by their distance to
the center, which is shown in column 4. Fields in different regions are
indicated by different font types: eastern fields in normal, western
fields in boldface and southern fields in italics. Mean metallicities are
very close to [Fe/H]$\sim$-1 in all fields within r$\lesssim$2$\fdg$5 from
the SMC center. A similar value is observed for the southern fields up to
r$\lesssim$3\arcdeg\@ (qj0047 and smc0049). For the outermost fields,
qj0033 in the West, and qj0102 and qj0053 in the South, the mean value is
clearly more metal-poor than in the others.

\begin{figure}[h!]
\begin{center} \includegraphics[width=4.2in]{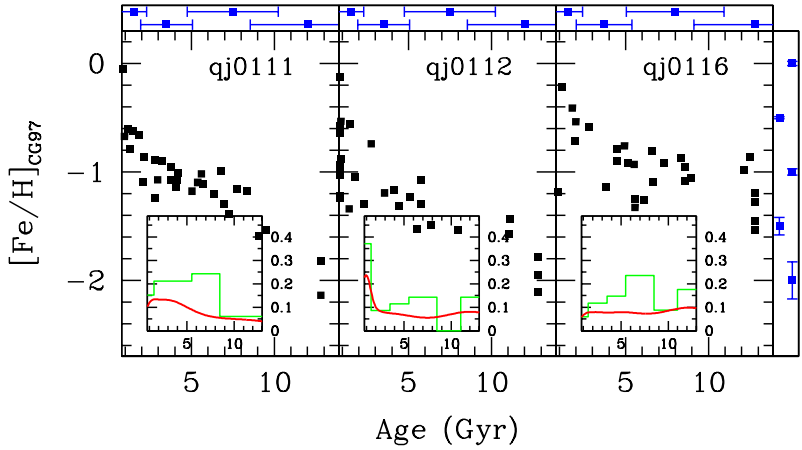} 
\caption{The same as Fig.\,\ref{fig2}, for the SMC eastern fields}
\label{fig4}
\end{center}
\end{figure}

The fact that the mean metallicity decreases when moving away from the
center implies that there is a metallicity gradient in the SMC. This is
the first time that a spectroscopic metallicity gradient has been reported
in SMC stellar populations. The detection of this gradient has been
possible because we have covered a large radius range, up to 4\arcdeg\@
from the SMC center.

\begin{figure}[h!]
\begin{center} \includegraphics[width=3.2in]{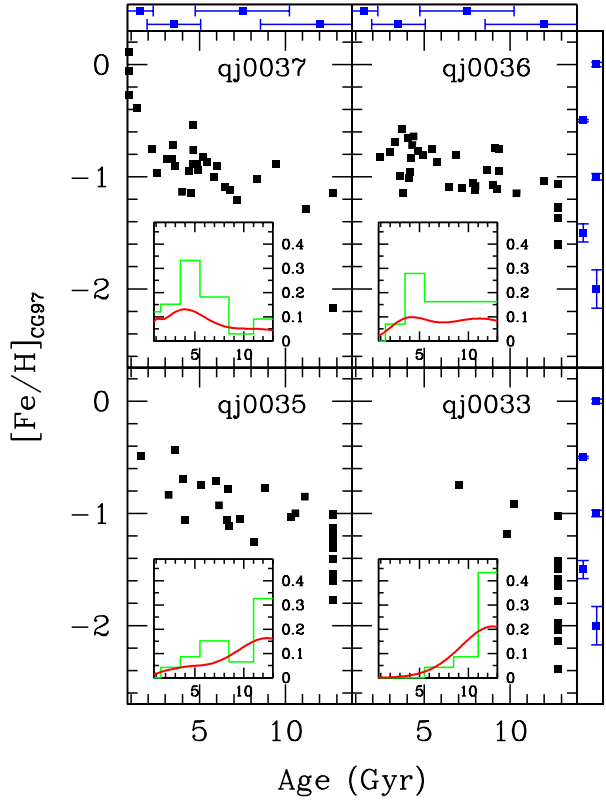} 
 \caption{The same as Fig.\,\ref{fig2}, for the SMC western fields}
\label{fig5}
\end{center}
\end{figure}

\begin{figure}[h!]
\begin{center} \includegraphics[width=4.2in]{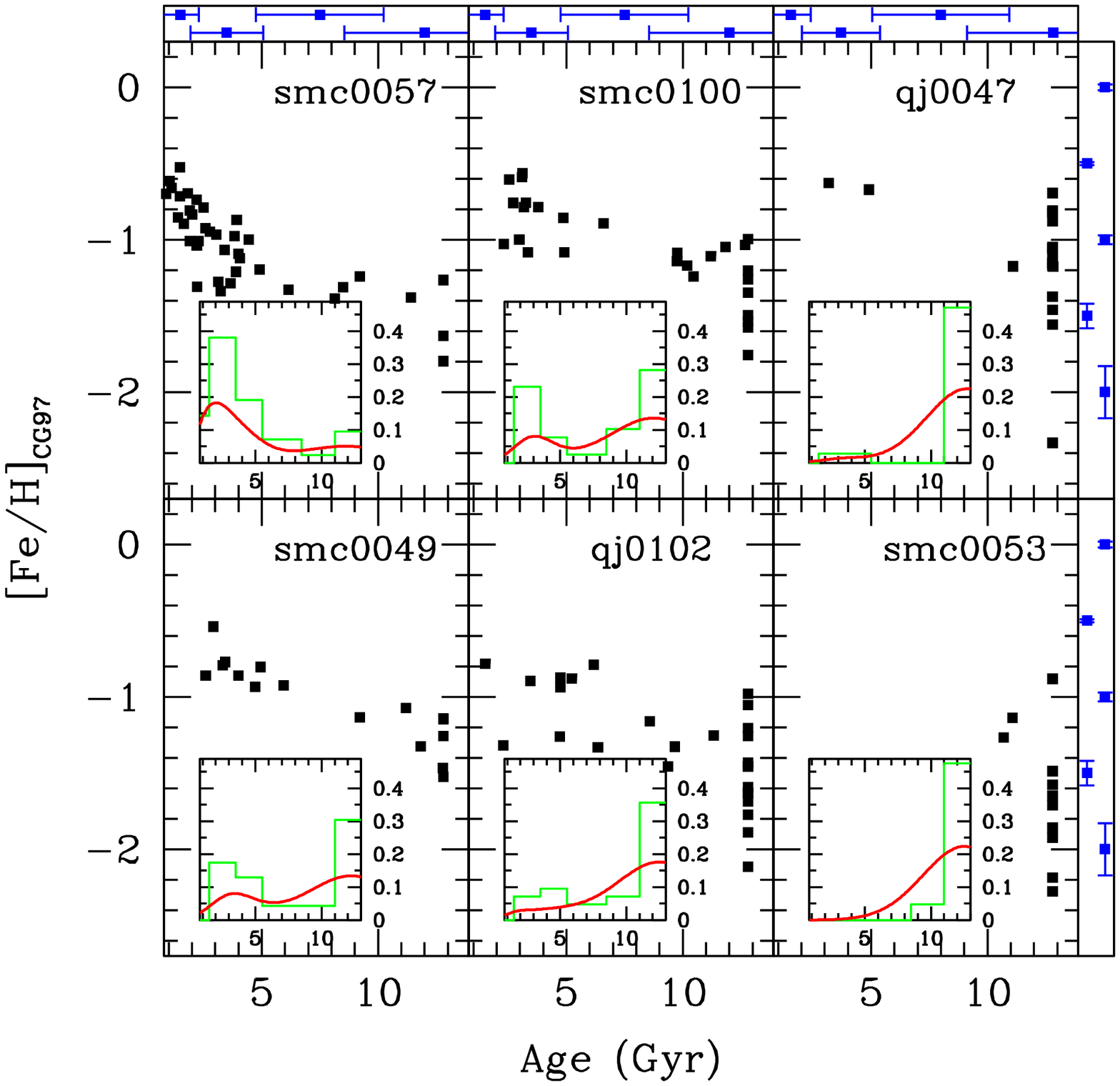} 
 \caption{The same as Fig.\,\ref{fig2}, for the SMC southern fields}
\label{fig6}
\end{center}
\end{figure}

To investigate the nature of the gradient, we have calculated
the AMR for each field. They are plotted in Fig.\,\ref{fig4},\, \ref{fig5}
and \ref{fig6} for fields situated to the East, West and South
respectively. Inset panels, as in Fig.\,\ref{fig2}, show the age
distribution of the observed stars, with and without taking into account
the age uncertainty (\textsl{solid line and histogram}, respectively).

All the AMR plotted in Fig.\,\ref{fig4},\, \ref{fig5} and \ref{fig6} show
a rapid chemical enrichment at a very early epoch. Even though in some
fields we have not observed enough old stars to sample this part of the
AMR, note that 12 Gyr ago all fields have reached [Fe/H]$\simeq -1.4-
-1.0$. This initial chemical enrichment was followed by a period of very
slow metallicity evolution until around 3 Gyr ago. Then, the galaxy
started another period of chemical enrichment, which is observed in the
innermost fields, which are, however, the only ones where we observed
enough young stars to sample this part of the AMR. In all cases, the mean
metallicity is similar to that of the other fields at similar
galactocentric radii. The field AMRs obtained in this work are similar to
those for clusters (the reader should take into account that differences
in the metallicity scales exist among different works), although there is
only one cluster older than 10 Gyr (e.g. \cite[Piatti \etal\ 2001]{piatti01}), and for planetary nebulae, with the
exception that in these objects it is not observed the chemical enrichment
episode at a very early epoch (\cite[Idiart \etal\ 2007]{idiart07}).

For eastern fields, located in the wing, most of the observed stars have
ages younger than 8 Gyr, but there is also a significant number of objects
older than 10 Gyr. At a given galactocentric distance, eastern fields show
a large number of young stars ($\leq$3 Gyr) in comparison to the western
ones. For the western and southern fields, the fraction of
intermediate-age stars, which are also more metal-rich, decreases as we
move away from the center, although the average metallicity in each age
bin is similar. This indicates the presence of an age gradient in the
galaxy, which may be the origin of the metallicity one. It is noticeable
that for the most external fields, qj0033 and qj0053, we find a
predominantly old and metal-poor stellar population.

\section{Future Work}\label{sec_conclusions}

We have investigated the chemical evolution of the field populations in
both Magellanic Clouds. However, there are still some points which should
be investigated. For example, in the LMC, we have studied 4 fields at
different galactocentric distances northward of the bar. Now it is
necessary to investigate what happens in other positions such as facing
the SMC and in the opposite direction. This could give us clues about the
effects of the interactions with the SMC in the LMC field populations.
Also, in both galaxies it is necessary to investigate the behaviour of the
abundances of different chemical species as a function of position in the
galaxy.

\end{document}